\documentclass[pdflatex,sn-nature]{sn-jnl}

\usepackage{graphicx}%
\usepackage{multirow}%
\usepackage{amsmath,amssymb,amsfonts}%
\usepackage{amsthm}%
\usepackage{mathrsfs}%
\usepackage[title]{appendix}%
\usepackage{xcolor}%
\usepackage{textcomp}%
\usepackage{manyfoot}%
\usepackage{booktabs}%
\usepackage{algorithm}%
\usepackage{algorithmicx}%
\usepackage{algpseudocode}%
\usepackage{listings}%
\usepackage{color, colortbl}

\theoremstyle{thmstyleone}%
\theoremstyle{thmstyletwo}%

\theoremstyle{thmstylethree}%

\usepackage{xr-hyper}   
\usepackage{hyperref}
\begin{document}

\title[Article Title]{An Energy-Efficient Adiabatic Capacitive Neural Network Chip}


\author[1]{\fnm{Himadri Singh} \sur{Raghav*}}\email{hraghav@ed.ac.uk}
\equalcont{These authors contributed equally to this work.}

\author[1]{\fnm{Sachin} \sur{Maheshwari}}\email{maheshwari.sachin@ed.ac.uk}
\equalcont{These authors contributed equally to this work.}

\author[1]{\fnm{Mike} \sur{Smart}}\email{msmart2@ed.ac.uk}
\equalcont{These authors contributed equally to this work.}

\author[2]{\fnm{Patrick} \sur{Foster}}\email{pfoster4@ed.ac.uk}

\author[1]{\fnm{Alex} \sur{Serb}}\email{aserb@ed.ac.uk}

\affil[1]{\orgdiv{School of Engineering, Institute for Integrated Micro and Nano Systems, Scottish Microelectronics Centre}, \orgname{University of Edinburgh}, \orgaddress{\city{Edinburgh}, \postcode{EH9 3FF}, \country{UK}}}
\affil[2]{\orgdiv{School of Engineering, Institute for Integrated Micro and Nano Systems, Centre for Electronics Frontiers}, \orgname{University of Edinburgh}, \orgaddress{\city{Edinburgh}, \postcode{EH9 3BF}, \country{UK}}}


\abstract{Recent advances in artificial intelligence, coupled with increasing data bandwidth requirements, in applications such as video processing and high-resolution sensing, have created a growing demand for high computational performance under stringent energy constraints, especially for battery-powered and edge devices. To address this, we present a mixed-signal adiabatic capacitive neural network chip, designed in a 130$nm$ CMOS technology, to demonstrate significant energy savings coupled with high image classification accuracy. Our dual-layer hardware chip, incorporating 16 single-cycle multiply-accumulate engines, can reliably distinguish between 4 classes of 8x8 1-bit images, with  classification results over 95\%, within 2.7\% of an equivalent software version. Energy measurements reveal average energy savings between 2.1x and 6.8x, compared to an equivalent CMOS capacitive implementation.}



\begingroup
\renewcommand{\thefootnote}{}
\footnotetext{This work has been submitted to \emph{Nature Communications} for possible publication. Copyright may be transferred without notice, after which this version may no longer be accessible.}
\endgroup

\maketitle

\clearpage
\section{Introduction} \label{sec:introduction}

Artificial Neural Networks (ANNs) were originally inspired by the structure and functionality of the human brain~\cite{Zhang2020,Rubino2021, Dalgaty2024}.
Today, ANNs built from artificial synapses and neurons are ubiquitous and serve as the computational backbone of modern AI systems, in a wide range of products and applications. As these ANNs have scaled, increasing computational and energy demands have put pressure on traditional digital compute platforms, such as SIMD accelerators, DSPs and GPUs. As such, more novel hardware implementations of artificial synapses have been explored, including CMOS subthreshold circuits~\cite{Yu2011, Rubino2021}, switched-capacitor approaches~\cite{Tsividis1987, Maundy1991, Hajtas2000, Noack2014, Ronchi2025}, oxide-based resistive RAM (OxRAM)~\cite{Garbin2015}, Li-ion transistor-based designs~\cite{Fuller2017}, memristors~\cite{Saxena2019, Papandroulidakis2019, Greatorex2025}, memcapacitors~\cite{Demasius2021, Pershin2014, Pershin2015, Wang2018}, and purely capacitor-based synapses~\cite{Cilingiroglu1991, Kwon2020, Hwang2022}, with the objective of increased energy efficiency.

Energy consumption reduction has obvious implications for battery life in mobile applications, the environment, cost and deployment infrastructures. Chip designers have sought ways to reduce energy through standard digital design techniques but these approaches have limitations~\cite{Nahmias2020}. Others have explored algorithmic and ANN model-level optimization, such as quantization or pruning~\cite{Sze2017}.
To that end, researchers have recently considered adiabatic capacitor-based synapse hardware that improves energy efficiency through slow charging, thereby avoiding current surges, intriguingly reducing dynamic energy dissipation significantly below the conventional discharging limit of $0.5\, CV^2$~\cite{Maheshwari2022, Massarotto2024}.
This efficient slow charging and discharging of capacitive synapses can be achieved using an adiabatic technique, also known as charge recovery logic~\cite{Athas1994, Koller1992, Younis1994,Teichmann2012}. 
Adiabatic operation mandates the use of a dedicated power supply that produces a sinusoidal, trapezoidal or step-charging waveform, known as a Power Clock (PC), which is capable of recycling energy, and is often implemented using either a capacitor-inductor resonator ~\cite{Maksimovid1995, Ziesler2001, Arsalan2005} or capacitor-capacitor step charging circuit ~\cite{ Raghav2016, Raghav2017, KhoramiA2020}. Adiabatic Logic (AL), which operates primarily in the analog domain, is thus very well-suited for low-power ANN applications~\cite{Ye1995, Blotti2004, Maheshwari2018, Massarotto2023, Raghav2020, Oklobdzija1997}.

\begin{figure}[t]
  \centering
  \includegraphics[width=\textwidth]{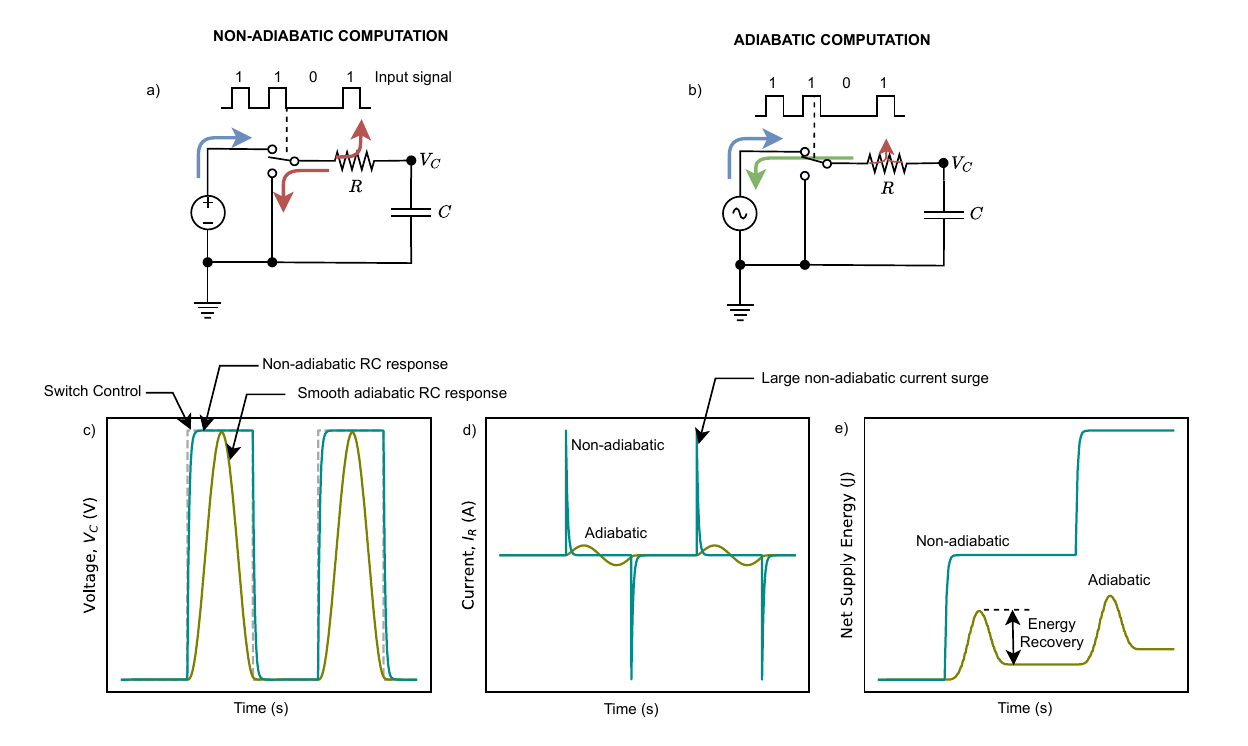}
  \caption{a) A standard switched non-adiabatic RC circuit, with a DC supply, where an input signal controls the switch state; b) An adiabatic RC circuit, with an inductive AC supply, where energy is recovered before the switch is grounded, leading to higher energy efficiency; c) Adiabatic and non-adiabatic $V_{C}$ node voltage response over two input signal cycles; d) Current response showing large current surges in the non-adiabatic circuit; e) Net energy response where adiabatic energy recovery and lower currents lead to significantly reduced net energy dissipation.}
    \label{fig:chip-architecture}
\end{figure}

Fig.~\ref{fig:chip-architecture}a shows the principle in action on a non-adiabatic circuit schematic: the capacitive load, $C$, is charged when the switch is closed, dissipating $0.5\, CV^2$ of energy. When the switch is reopened, the capacitor is discharged to ground, dissipating another $0.5\, CV^2$, giving a total dissipation of $CV^2$. In the adiabatic circuit of Fig.~\ref{fig:chip-architecture}b, the capacitive load is charged by a sinusoidal PC supply~\cite{Maksimovic2000,Ziesler2001, Maheshwari2022} during the PC upswing (evaluation phase) and discharged back into the supply during the downswing (recovery phase), recycling energy that would otherwise be dissipated to ground. The gradual ramping of the PC both prevents current surges and also reduces resistive losses, as shown in Fig.~\ref{fig:chip-architecture}c-d. The resulting differences in the net supply energy over time, for adiabatic versus non-adiabatic circuits, are shown in Fig.~\ref{fig:chip-architecture}e.

In this paper, we present the physical implementation of a 64-bit input, dual-layer, adiabatic neural network image classification chip and show measured energy and accuracy results. The chip implements a hard-coded ANN inference architecture including 16 mixed-signal processors, running at 1$MHz$, each capable of single-cycle Multiply-Accumulate (MAC) computations. The choice of a hard-coded pipeline maximizes energy efficiency by removing configuration switching overheads, and targets heavy-duty pretrained applications, such as front-end feature extraction backbones. 
The chip was fabricated in a commercially available 130$nm$ CMOS technology using custom Metal-Oxide-Metal (MOM) capacitor arrays and mounted on a custom PCB with two integrated adiabatic Power Clock Generator (PCG) power supplies to drive computations.
Using single-bit output neurons, the chip reliably classifies 8×8 binary images into four classes with $>95\%$; accuracy and a $<$2.7$\%$ deviation from software ANN performance. Operating at 1$MHz$ and 1.5$V$, it achieves 2.1× to 6.8× reductions in energy dissipation relative to an equivalent CMOS-capacitive design.

\subsection{The Adiabatic Neuron: The Computational Workhorse}\label{sec:single-neuron-design}

\begin{figure}[t]
  \centering
  \includegraphics[width=\textwidth]{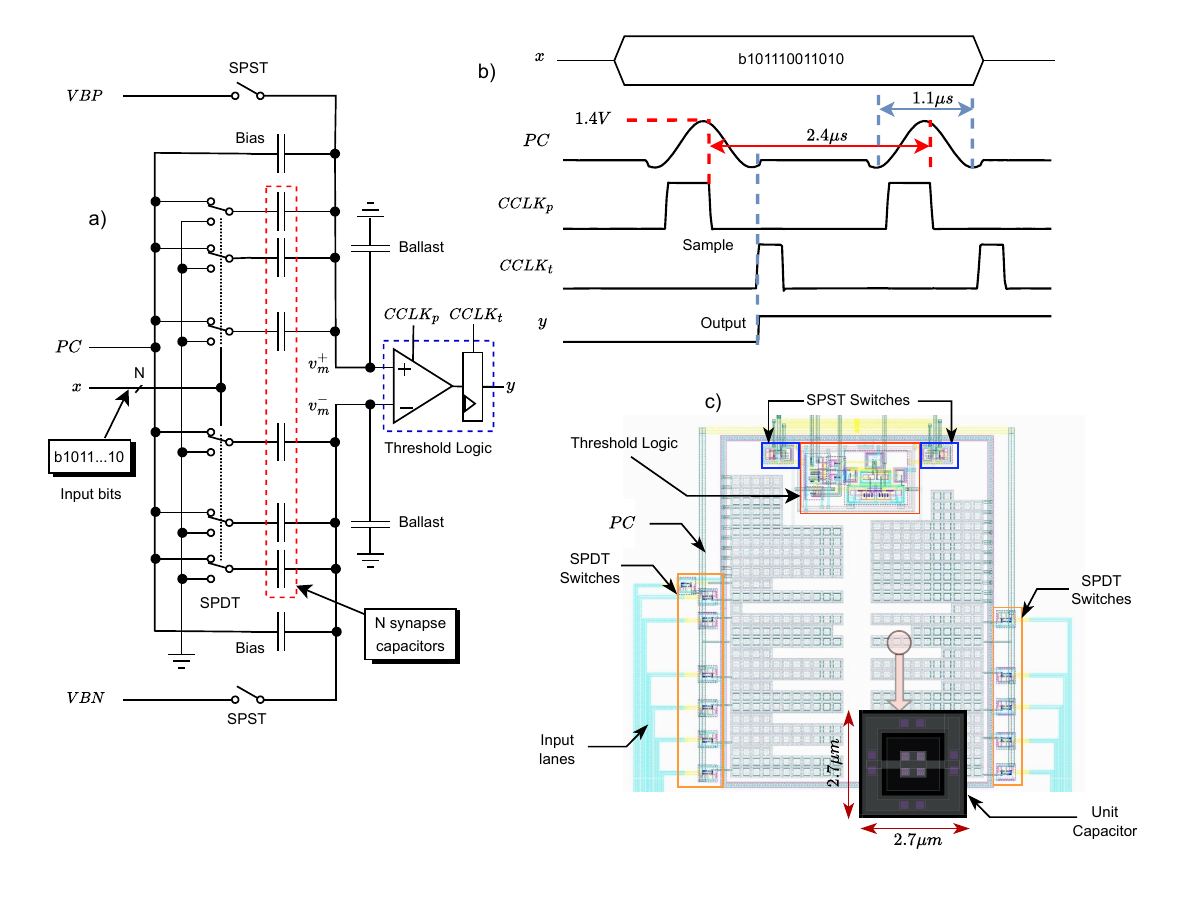}
    \caption{The ACN computational workhorse: a) Schematic of a DTSC ACN with a single power clock signal $PC$, an $N$-bit input vector $x$, $N$ synapse capacitors, threshold logic unit and ACN output, $y$; b) Experimentally measured oscilloscope trace of the primary ACN signals illustrating operational timings of one of the 12 \mbox{1-bit} input ACNs in our chip; c) 130$nm$ CMOS layout of the same ACN.}
    \label{fig:single_acn}
\end{figure}

Artificial neural networks typically consist of multiple layers of computational units, densely linked with weighted interconnections. The Artificial Neuron (AN) serves as the core computational unit within each ANN. Each AN is required to perform a large number of energy-intensive MAC operations every second, combining input signals with abstract weight, or \emph{synapse}, values before applying a nonlinear activation function. To address the significant power demands of work-intensive MAC operations, we see adiabatic capacitive hardware offering a promising approach to drastically reduce energy consumption, even with minor trade-offs in computational accuracy due to the shift from traditional digital to analog processing. In our chip, the Adiabatic Capacitive Neuron (ACN), introduced in previous work~\cite{Maheshwari2025} and reproduced in Fig.\ref{fig:single_acn}a, serves as a functionally-equivalent, energy-efficient, computational AN workhorse.

A prominent feature of the ACN is its dual switched-capacitor arrays, or \emph{trees}, representing the positive and negative-weighted synapses, respectively. Each of the $N$ synapse capacitors is connected to the sinusoidal PC dependent on a switch controlled by the corresponding bit of the $N$-bit ACN input, $x$; otherwise, the capacitor remains grounded. 
The \emph{membrane} nodes, $v_{m}^{\pm}$, then perform multi-input MAC operations via the principle of capacitive charge redistribution and voltage division.
Bias capacitors, ballast capacitors and the connections to $VBN$ and $VBP$ DC voltages enable the implementation of neuron biases, $v_{m}^{\pm}$ node swing moderation and periodic $v_{m}^{\pm}$ node resetting, respectively. Detailed explanations can be found in~\cite{Smart2025}.

The Threshold Logic (TL) unit implements the AN activation function by comparing the membrane voltages formed at the TL inputs and generating a binary output, $y$. 
As shown in Fig.~\ref{fig:single_acn}b, if $v_{m}^{+}>v_{m}^{-}$ at the time of sampling defined by the $CCLK_{p}$ signal, the TL latches a binary output value of 1 on the rising edge of $CCLK_{t}$, otherwise, 0. Like the PCG, these non-overlapping clock signals are also generated externally on the PCB. The TL is powered by a non-adiabatic DC supply. However, the high impedance TL unit plays a vital role in adiabatic operation. It allows the capacitive energy stored during computation to be recovered by the PCG, and importantly, it allows small differential membrane voltages to control ACN switching in subsequent adiabatic layers.

An example of the physical layout of one of the ACNs in the chip, including the placement of its various modules, is shown in Fig.~\ref{fig:single_acn}c. 
The synapse capacitors in the two trees use custom MOM capacitors designed specifically for this architecture. Fig.\ref{fig:single_acn}c shows an inset of the square-slotted unit capacitor of $\sim2fF$ having a square area of $2.7~\mu\text{m}\times2.7~\mu\text{m}$. The slots in the metal layers aid in constructing larger-valued capacitors, thus reducing thermal stress during fabrication. These larger ACN capacitance values were created by interconnecting multiple unit capacitors.

\subsection{An ACN Network}
\label{sec:chip-architecture}

\begin{figure}[t]
  \centering
  \includegraphics[width=\textwidth]{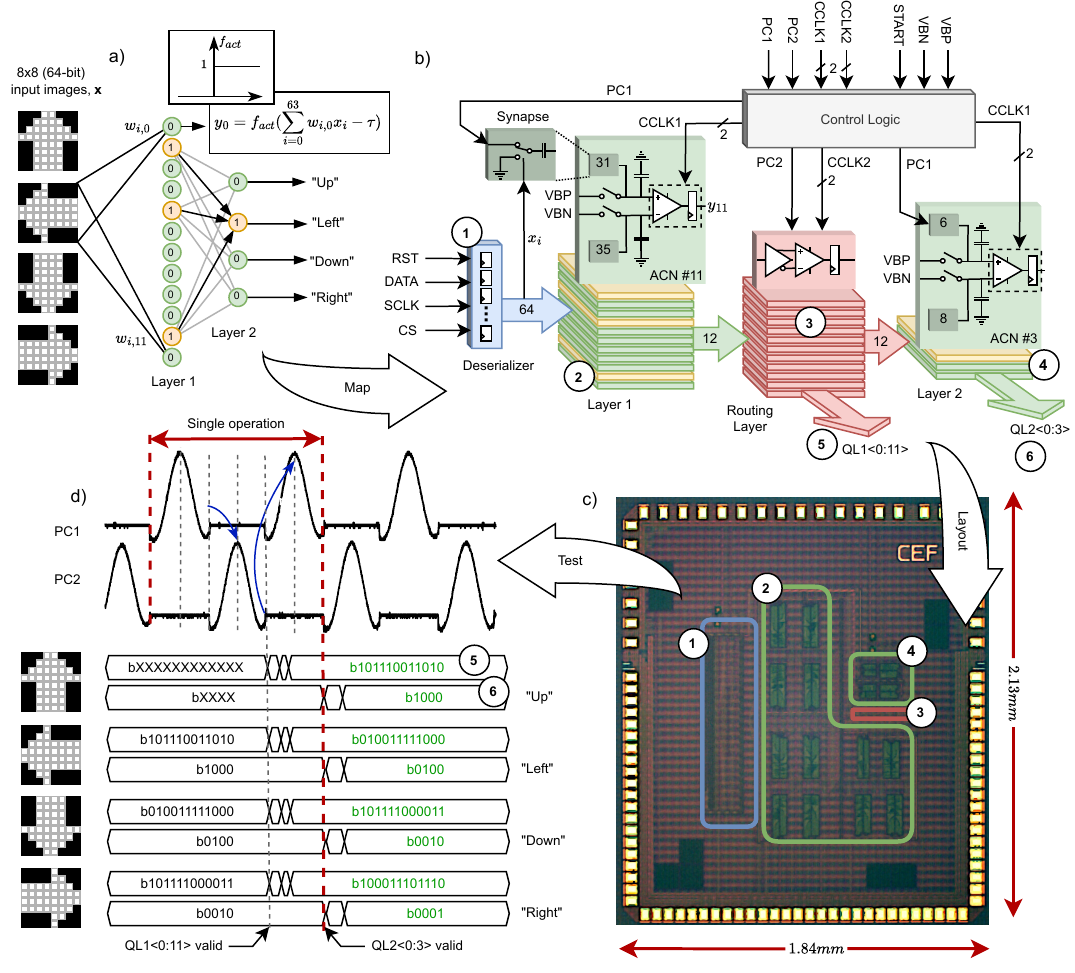}
    \caption{End-to-end adiabatic image classification: a) 64x12x4 TensorFlow-trained software ANN arrow classifier with 64 1-bit inputs $x$, weights $w$, software bias $\tau$ and Heaviside activation function, $f_{act}$; b) chip architecture block diagram of mapped ACNN implementation; c) contrast-enhanced micro-photograph of the ACNN 130$nm$ chip; d) stacked oscilloscope traces showing ACNN outputs across 4 independent functional test operations using $PC1$ for the ACN layers and a separate $PC2$ for the routing layer. Blue arrows represent data transitions between layers.}
    \label{fig:end_to_end_chip}
\end{figure}

We can now introduce the Adiabatic Capacitive Neural Network (ACNN) as an efficient hardware implementation of a software-trained ANN, based on the highly energy-efficient capacitive ACN discussed in the previous section. To demonstrate that an ACNN was physically realizable using multiple ACNs, distributed across multiple layers, we designed a fully operational ACNN chip. Our chip was designed to work as an image classifier.
Demonstrating that our ACNN chip could fully implement an inference engine derived from a TensorFlow-trained~\cite{Abadi2016} software neural network, using real-world data, was a key objective. Equally crucial was validating that this ANN inference capability could be transferred to the ACNN with negligible loss in classification accuracy, while enabling energy-efficient computation.

Fig.~\ref{fig:end_to_end_chip} shows our end-to-end process from trained software ANN, through circuit design, physical chip and finally testing of the ACNN with equivalent input data. To start the process, we selected a two-layer software ANN, as shown in Fig.~\ref{fig:end_to_end_chip}a, with 12 hidden layer neurons and 4 outputs. All 16 neurons used an ACN-compatible Heaviside activation function to binarize the output from each node in the network. The ANN was trained on over 4000 samples from the 8x8 pixel, 1-bit \emph{arrows8} dataset~\cite{arrows2025} with 98.65\% accuracy. Details of the actual training process are provided in Method \ref{sec:method_ann_training}. Each image in the dataset is uniquely classified into one of four categories: \emph{Up}, \emph{Left}, \emph{Down}, and \emph{Right}.

Each neuron, across the two layers, in the trained software ANN was optimally mapped to capacitance values to generate 16 unique and hard-coded ACN designs.
Due to the analog nature of the ACN computation, it was possible that the ACNN might not match exactly with the original 98.65\% software ANN performance. As such, quantifying any potential losses was a key objective. 
One reason was that although the preliminary mapping of software weights to ACN capacitance values is a lossless process~\cite{Smart2025}, the instantiated, physically realized ACN capacitance values would differ in size from the required mapped value. The use of the $\sim$2$fF$ unit MOM capacitor was one source of quantization error. Parasitic-to-ground capacitances on the $v_{m}$ nodes generated by the capacitive array, normally an unwanted design feature, are neatly used to generate some of the required ballast capacitance, using a simple compensation process. Moreover, using predicted post-layout capacitance values, with an average predicted error of 0.53$fF$, an ACNN software model estimated a drop of 0.39\% to 98.26\%. Details of the mapping, model and ACNN network analysis can be found in Extended Data Fig.~\ref{extfig:to2_config} and Method~\ref{sec:method_ann_mapping}.

Fig.~\ref{fig:end_to_end_chip}b shows the architectural block diagram of our complete ACNN chip. The 64-bit test images are clocked in via a 1$MHz$ SPI serial interface to the ACNN \emph{Deserializer} shift register. The resulting 64-bit wide parallel input signal is then provided to the first layer of 12 mapped and independently instantiated ACNs. 
The outputs generated by the first layer are routed to a second layer of 4 ACNs through an intermediate routing layer. The 4-bit inference output, $QL2$, is subsequently transmitted to the external chip pads for off-chip readout. The 12-bit outputs from our first layer are also routed to external chip pads as $QL1$, for debugging purposes. The two ACN layers represent about 51$pF$ of distributed capacitance, driven by an externally provided single power clock, $PC1$, via the control logic, which also provides the two associated non-overlapping TL clock signals, $CCLK1$ and $CCLK2$. A description and purpose of the other control logic signals in our chip are provided in Method~\ref{sec:method_control_logic}.

To address timing constraints in the multi-layer ACNN, a dedicated Routing Layer (RL) is sandwiched between consecutive computational layers. This layer decouples the two ACN layers, regulates signal transfer, mitigates delay variations, and maintains synchronization throughout the system.
As seen in Fig.~\ref{fig:end_to_end_chip}b, the RL comprises an adiabatic buffer, a Dynamic Latch Clocked Comparator (DLCC) circuit, followed by a clocked SR latch stage. The adiabatic buffer generates both true and complementary output signals. The true signal is fed to the positive input terminal of the DLCC. In contrast, the complementary signal is fed to the negative input terminal. The adiabatic buffer is driven by a second independent external power clock, $PC2$, with associated $CCLK2$ signals used in the DLCC and latch. The $PC2$ is 180$^\circ$ out of phase with the $PC1$ used for the ACN layers. Finally, the latch output, acting as a control input, is fed to the second-layer switches. 

A total of 10 ACNN chips were fabricated. A micrograph of one of the 1.84 mm x 2.13 mm chips is shown in Fig.~\ref{fig:end_to_end_chip}c, highlighting peripherals, such as the \emph{Deserializer} logic, the two neural ACN layers and the connecting routing layer. The ACNs and RL, the computational heart of the chip, occupy an internal core area of 1145$\mu m$ x 1307$\mu m$. 

The last stage of the end-to-end process was chip validation, the results of which we will explore in the following sections. The test images were applied to the ACNN chip, one sample at a time. The first layer computation, performed on the first $PC1$ peak, is passed to the RL after this initial pulse. This intermediate result is then latched and made available to the second layer after the next $PC2$. Finally, the ACNN generates the final ACNN output on the second pulse $PC1$. This multi-cycle, input-to-output, classification process is defined as a single computational \emph{operation}.
Fig.~\ref{fig:end_to_end_chip}d shows oscilloscope traces from four independent test operations processed one after another, stacked vertically for readability.
A chip is functioning correctly if $QL2$ matches exactly with the original ANN software, given the same input test pattern.

\subsection{An Energy-Efficient ACNN Image Classifier}\label{sec:measurements}

The fabricated ACNN chips allowed us to verify whether the ACNN was indeed capable of energy-efficient, accurate and reliable computation. To test the ACNN chips, we used an ArC TWO (ArC2) instrumentation board~\cite{Arc2025}. This required the development of a custom daughter board to provide the interface between the ArC2 and the ACNN Device Under Test (DUT) chip. The daughter board incorporated both the PCG circuitry and TL clock signal generators. The oscilloscope traces in Figs.~\ref{fig:single_acn},\ref{fig:end_to_end_chip} show the non-ideal nature of power clock generation from real power supply circuits, such as the one implemented on our custom daughter board. The operation of the PCG circuitry on the PCB, and associated non-idealities such as the switched capacitive load presented by the DUT, are discussed in Method~\ref{sec:PCG}. Extended Data Fig.~\ref{extfig:pcg} provides an overview of the working and operation.
An overview of the entire test setup is shown in Extended Data Fig.~\ref{extfig:test_set_up} and the operation is described in more detail in Method~\ref{sec:method_test_procedure}.

\subsubsection{Cross-Chip Classification}\label{sec:cross-chip-classification}

For functional validation, we applied the complete set of 4,078 \emph{arrows8} test samples to the ACNN chip and recorded the resulting outputs.
Before each image classification operation, the LC-based PCG tank capacitors were fully topped up to avoid any PC magnitude decay over the full test dataset duration. To account for variations from non-ideal analog computation and process-dependent factors, such as device mismatch, we repeated the experimental measurements across five randomly selected ACNN chips, with each chip characterized over 10 repeated iterations to capture intra-chip variability.

\begin{figure}[t]
  \centering
  \includegraphics[width=\textwidth]{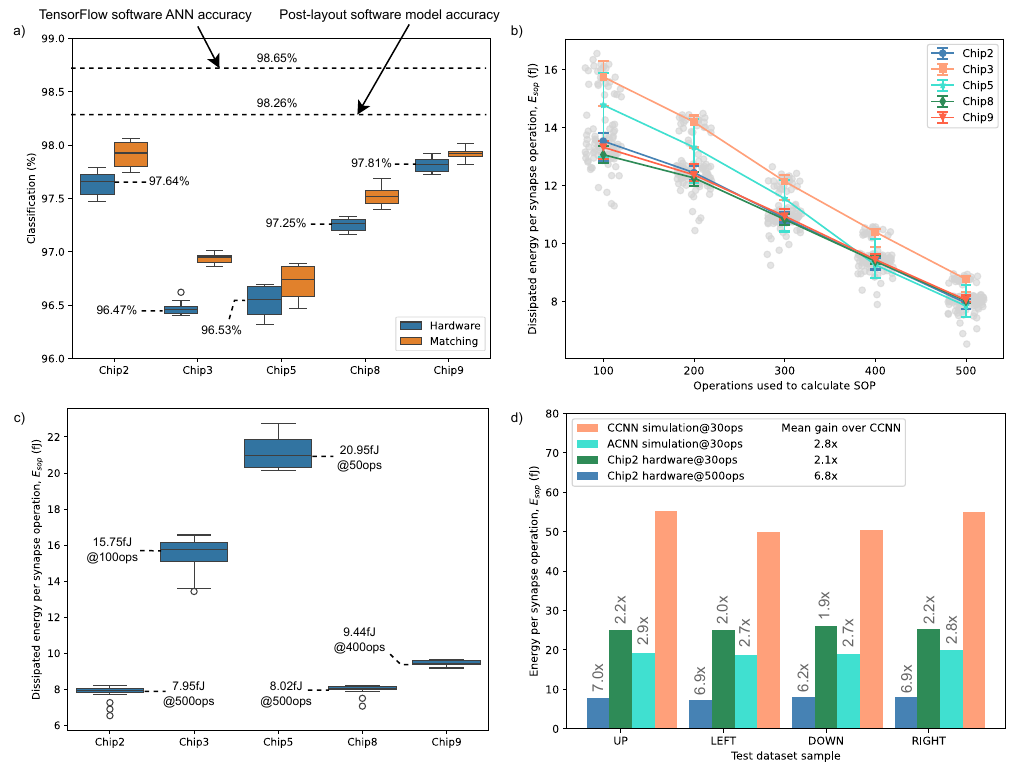}
    \caption{ACNN chip results: a) Cross-chip classification results as box-and-whisker plots showing median, InterQuantile Range (IQR), with whiskers at 1.5IQR. One outlier is shown for \emph{Chip3}. Hardware results annotated with sample mean; b) Cross-chip $PC1$ $E_{SOP}$ median energy results with 1.5IQR error bars. Energy calculated over different numbers of operations, across 4 test images and 5 iterations per test, with jittered grey operation samples to aid visualization; c) Boxplot of dissipated SOP chip energy after several operations, without significant classification errors, using 4 data samples and 5 test iterations per sample. Chip result annotated with median energy and number of operations (ops); d) Comparison of hardware versus ACNN post-layout and CCNN schematic simulations.}
    \label{fig:to2-results}
\end{figure}

The silicon measurement results are provided in Fig.~\ref{fig:to2-results}a, where \emph{Hardware} refers to the absolute ACNN hardware classification rate, and \emph{Matching} quantifies the degree to which software and hardware outputs coincide, regardless of correct image identification. The results show that the performance of the hardware classification does vary slightly from iteration-to-iteration for each chip, not unexpectedly due to the analog nature of the computation, but with a standard deviation of less than 0.14$\%$, indicating high reliability. A \emph{Shapiro-Wilk} test ($p>0.05$) indicates possible normality in the results per chip, but this is based on only 10 samples. The error between the mean hardware and the original TensorFlow software ANN, across the five chips, varies between $0.84\%$ and $2.18\%$, with $0.39\%$ of that error predicted by software, coming from the layout and parasitic compensation process. Chip-to-chip variations are associated with manufacturing issues, such as process speed and mismatch, that affect analog computation.

To understand the source of the small and slightly varying number of extra hardware misclassifications in Fig~\ref{fig:to2-results}a, a deeper analysis of the results was required.
We observed good supporting evidence that the internal ACN membrane voltage differentials $\Delta v_{m}$, used for the computation inside our chip, were close to the values predicted by simulation, with an average error of around 18$mV$. 
A further analysis of the hardware classification results determined that across all tests and chips 90.3\% of the test samples were always classified identically to the ANN. We examined the predicted properties of the remaining mismatching 9.7\% samples using the software ACNN model. Whilst there was a slight propensity towards the \emph{Down} class, generally there was nothing obviously different between these samples, compared to the entire test dataset (Extended Data Fig.~\ref{extfig:to2_class_errors}a,b). This is until we look at the predicted absolute TL input differential voltages, $|\Delta v_{m}|$. In this case, the mismatching samples have, in general, much smaller values (Extended Data Fig~\ref{extfig:to2_class_errors}c,d). If we look specifically at the ACNs that produced a bit error in the hardware, we see predicted $|\Delta v_{m}|$ very close to zero, with 89\% of samples below 30$mV$. This is a clear, practical demonstration of the importance of differential input voltage to overall performance, as previously suggested theoretically in~\cite{Smart2025}. Extended Data Fig.~\ref{extfig:to2_class_errors}e,f shows the differences between the two ACNN layers. From the predicted $|\Delta v_{m}|$, it was evident that layer 1 had the greater potential for any errors, as reflected by the hardware bit error rate. The reason for the robustness of the network classification in our chip, even when using analog computation, is plain to see with the values predicted by layer 2 $|\Delta v_{m}|$ predominantly above 100$mV$.

\subsubsection{Cross-Chip Energy Efficiency}\label{sec:energy-testing}
The ACNN energy efficiency chip testing focused on the $PC1$ supply that powers all ACN synapse computations. To calculate the dissipated energy per synapse operation, $E_{SOP}$, we classified a test image over multiple operations, \emph{without} topping up the $PC1$ tank capacitor, and measured the subsequent voltage drop across this capacitor. For each chip, we tested with 5 different numbers of successive operations 
using 4 different test samples: UP, LEFT, DOWN and RIGHT, as defined in the Table~\ref{tab:arrows8_specific_samples}. 
Each test was repeated over 5 iterations to determine variance. 
Extended Data Fig.~\ref{extfig:energy_analysis}a
shows the voltage drop on the $PC1$ tank capacitor, measured after different numbers of consecutive operations.
The resulting computed total energy dissipated, which includes the PCG, is provided in Extended Data Fig.~\ref{extfig:energy_analysis}b, with the case with no DUT representing the PCG energy contribution, $E_{PCG}$. For a detailed description of the methodology for measuring energy and a formal definition of $E_{SOP}$, see Method~\ref{sec:method_measure_energy}.

\begin{table}[h]
\caption{Specific 8x8 1-bit test samples from the \emph{arrows8} dataset test split used in this paper, with zero-indexed split indices.}
\label{tab:arrows8_specific_samples}%
\begin{tabular}{@{}llll@{}}
\toprule
Sample Name & Split & Class & Sample Index \\
\midrule
UP & test & \emph{Up} & 102 \\
LEFT & test & \emph{Left} & 70 \\
DOWN & test & \emph{Down} & 48 \\
RIGHT & test & \emph{Right} & 23 \\
\botrule
\end{tabular}
\end{table}

Fig.~\ref{fig:to2-results}b presents the chip-to-chip $E_{SOP}$ figures, with grey jittered data points corresponding to the full set of experimental measurements. The median and quartile estimates show a pronounced central clustering of energy values with few outliers, with the narrow quartile intervals of the data highlighting the robustness and repeatability of the experimental observations. With increasing numbers of operations, and larger measured voltage drops, Fig.~\ref{fig:to2-results}b shows reducing deviations in the results, suggesting enhanced experimental reliability.
However, most intriguingly, the graph also reveals a clear trend in which increasing the number of operations used to compute $E_{SOP}$ leads to a systematic decrease in energy dissipation, with the lowest consumption observed at 500 operations. 
Sustained operation without recharging the tank capacitor between operations, though, does lead to reduced peak $PC1$ voltages, $V_{max}$, over time, as shown in Extended Data Fig.~\ref{extfig:energy_analysis}c. With lower voltages comes lower energy dissipation~\cite{Maheshwari2025}.

The voltage scaling properties of ACNs~\cite{Smart2025} mean that, theoretically, the computation should not be affected by reduced PC peak voltages. However, this natural and exceptionally useful benefit of the ACN has practical limitations, with reduced differential voltages being more susceptible to TL-based errors. Extended Data Fig.\ref{extfig:energy_analysis}d shows that as the number of energy operations increased, so did the error rate. Fig.~\ref{fig:to2-results}c shows a box-and-whisker $E_{SOP}$ plot on our five test chips. Notably, the $E_{SOP}$ values presented here are taken at the number of operations before the mis-classification rate exceeds 5\%. The results highlight noticeable differences in the energy distribution between devices.

The energy measurements show strong test repeatability for each chip individually, with moderate chip-to-chip variation becoming apparent as the operation count increases. Across the measured devices, \emph{Chip3} and \emph{Chip5} are stable within ~70-100 operations (ops), while others, including \emph{Chip2}, \emph{Chip8}, and \emph{Chip9}, remain robust over a substantially broader operational window of up to~500 ops. The differences observed across all devices are well aligned with expected process and device-level variability, which naturally leads to variation in stability, dissipation, and performance consistency.

To verify the ACNN chip measurements, a comparative analysis was performed against energy estimates derived from post-layout simulations. For consistency, the simulations employed the same non-ideal PCG circuit as implemented on the daughter board. 
The simulated $E_{SOP}$ values are approximately $\sim 5fJ$ less from the silicon measurements, demonstrating strong agreement and validating both our energy measurement methodology and modeling framework.
Using the same four test samples as before,
Figure~\ref{fig:to2-results}d presents a $E_{SOP}$ comparison at 30 operations between \emph{Chip2}, post-layout ACNN and its counterpart CMOS Capacitive Neural Network (CCNN) powered by a DC supply. 
At 30 operations, the results show that \emph{Chip2} and post-layout ACNN achieve an average 2.1× and 2.8× reduction in energy dissipation, respectively, compared to the simulated CCNN. 
At 500 operations, the results demonstrate that the silicon-tested ACNN design achieves further significant energy savings of an average 6.8×, relative to the CMOS baseline.
The total energy dissipation for the post-layout ACNN and the pre-layout CCNN is presented in Table~\ref{tab:ACNN_Energy} and Table~\ref{tab:CCNN_Energy}, respectively.

\begin{table}[h]
\caption{The total $PC1$ dissipated energy of ACNN post-layout at the nominal corner under a 1$MHz$ resonant power-clock. To isolate overhead contributions, results are provided for both the synaptic circuitry alone and the combined system, including PCG energy. The reported values are presented for 1, 10, and 30 multiple operations.}
\label{tab:ACNN_Energy}%
\begin{tabular}{@{}lllllll@{}}
\toprule
   & \multicolumn{6}{c}{\textbf{Total ACNN Energy} ($pJ$)} \\
\hline
   & \multicolumn{3}{c}{\textbf{With PCG$^a$}} &\multicolumn{3}{c}{\textbf{Without PCG$^b$}} \\
\hline
Sample Name & 1 & 10 & 30 & 1 & 10 & 30 \\
\midrule
UP & 25.03 & 247.98 & 729.06 & 16.46 & 162.32 & 472.08 \\
LEFT & 24.47 & 242.47 & 712.99 & 15.91 & 156.80 & 456.01 \\
DOWN & 24.73 & 245.02 & 720.56 & 16.17 & 159.36 & 463.58 \\
RIGHT & 25.60 & 253.42 & 744.88 & 17.10 & 168.22 & 489.04 \\
\botrule
\end{tabular}
 \begin{tablenotes}
\footnotesize
\item[$^{a}$]This is the total post-layout energy to the number of energy operations (1, 10, 30), denoted as $E_{\text{total}}(O_{\text{max}})$, which accounts for the energy dissipated by the PCG in driving the ACNN chip.
\item[$^{b}$]This is the total post-layout energy dissipated, excluding the PCG energy. To calculate $E_{SOP}$, the energy values are divided by the number of synapses and operations, $N_{\text{synapses}} \times O_{\text{max}}$.
    \end{tablenotes}
\end{table}

\begin{table}[h]
\caption{The total pre-layout dissipated energy of CCNN under nominal corner at 1$MHz$ clock. The reported values are presented for 1, 10, and 30 multiple operations.}
\label{tab:CCNN_Energy}%
\begin{tabular}{@{}llll@{}}
\toprule
   & \multicolumn{3}{c}{\textbf{Total CCNN Energy ($pJ$)}} \\
\hline
Sample Name & 1 & 10 & 30  \\
\midrule
UP & 46.02 & 455.30 & 1352.65  \\
LEFT & 40.74 & 402.86 & 1198.15  \\
DOWN & 41.93 & 414.99 & 1231.21  \\
RIGHT & 45.87 & 453.59 & 1347.47  \\
\botrule
\end{tabular}
\end{table}


\begin{sidewaystable}
\caption{Comparison of this work with state-of-the-art.}
\label{tab:comparison}%
\begin{tabular}{@{}lllllll@{}}
\toprule
References & ISSCC'20~\cite{Wan2020} & SSCL'21~\cite{Agrawal2021} & TCAS-I'22~\cite{Maheshwari2022} & npj'24~\cite{Massarotto2024} & NCE'24~\cite{Richter2024} & This Work$^{**}$\\
\midrule
\rowcolor{brown!40}
Technology & 130$nm$ & 65$nm$ & 180$nm$ & 180$nm$ & 180$nm$ & 130$nm$ \\
\rowcolor{brown!20}
Implementation &  Mixed-Signal & Digital &  Mixed-Signal & Mixed-Signal & Asyn Mixed-Signal & Mixed-Signal  \\
\rowcolor{brown!20}
 &  CMOS/RRAM  & CIM & Adiabatic & Adiabatic & Subthreshold & Adiabatic  \\
 \rowcolor{brown!40}
Supply Voltage & 1.8V & 0.85V  & 1.8V & 1.8V & 1.8V & 1.5V \\
\rowcolor{brown!20}
Synapse Type & RRAM & 10T-SRAM & 1-bit MIM & 8-bit BEOL & Dynamic Analog & 1-bit MOM \\
\rowcolor{brown!20}
 &  &  & Capacitors & Capacitors & Current-mode Circuit & Capacitors  \\
\rowcolor{brown!40}
Neuron Type &  IF  & IF/LIF/RMP &  Comparator & IF & LIF/exLIF & Comparator  \\
\rowcolor{brown!20}
\# of Neurons &  256 & 192 & 1 & 256 & 1024 & 16  \\
\rowcolor{brown!40}
\# of Synapses &  65k & 1.5k  & 4 & 65k & 65k & 816  \\
\rowcolor{brown!20}
\# of Layers & -- & --  & 1 & -- & -- & 2  \\
\rowcolor{brown!40}
Core Area & 1.79 $mm^2$ & 0.089 $mm^2$ & -- & -- & 69.94 $mm^2$ & 1.497 $mm^2$ \\
\rowcolor{brown!20}
Energy & 0.0139fJ/MAC$^a$ & 1.01pJ/op$^b$ & 0.04725pJ/op$^c$ & 0.47pJ/op$^d$ & 150pJ$^e$ & 0.00778pJ \\
\rowcolor{brown!20}
& & \emph{@}200$MHz$ & \emph{@}1$MHz$ & \emph{@}0.5$MHz$ & \emph{@}80$Hz$ & \emph{@}1$MHz$ \\
\rowcolor{brown!20}
& (Experimental) & (Experimental) & (Simulation) & (Simulation) & (Experimental) & (Experimental)  \\
\botrule
\end{tabular}
 \begin{tablenotes}
\footnotesize
\item[$^{**}$]Here the core area includes peripheral circuitry. \emph{Chip2} $E_{SOP}$, excluding the TL energy, is given at 500 operations.
\item[$^a$]Here, 1 MAC counted as 2 operations.
\item[$^b$]Energy for the main synaptic operation, the AccW2V instruction. Here, 1 op is equal to a 11-bit signed operation.
\item[$^c$]This is the pre-layout energy per synapse operation ($E_{SOP}$) for a 1-bit synapse operation, excluding TL energy.
\item[$^d$]Pre-layout energy per synaptic operation, including neuron energy.
\item[$^e$]Energy per “Neuron Spike Operation", comprising one-to-many MAC-like operations depending on the integration period and on the assumed time-step fidelity.
    \end{tablenotes}
\end{sidewaystable}

In Table~\ref{tab:comparison}, we compare the performance of our energy-efficient ACNN with previously published work that includes SRAM, RRAM, capacitive and subthreshold implementations of the synapse. Our chip prioritises energy efficiency by using capacitive synapses in adiabatic mode rather than a non-adiabatic CMOS mode. Mixed-signal RRAM/CMOS systems tend to reach slightly better energy efficiency compared to non-adiabatic CMOS implementations and higher chip density, albeit at the cost of high RRAM forming voltages. 
This is the first in-silico demonstrated adiabatic neural-network chip to deliver competitive energy performance vs non-adiabatic hardware systems as seen in table \ref{tab:comparison} while simultaneously achieving software-level classification accuracy.

\section{Discussion}\label{sec:discussion}

This paper introduced a highly energy-efficient, adiabatic neural network chip capable of performing image classification. At the heart of the chip are the single-cycle, differential, multiply-accumulate engines, based on a minimal, custom MOM, switched-capacitor array structure, all driven by an energy recycling supply. These computational engines, with inherent natural resilience to non-idealities, combine in multiple numbers and
interconnected layers to form a robust network capable of accurate image classification. We have demonstrated the ability to build such a chip, end-to-end, from training a software neural network in a standard AI framework and optimally mapping it to a fully functioning and reliable inference chip.

Our ACNN shows robust, reproducible performance across multiple fabricated chips. Comprehensive validation, including functionality, cross-chip consistency, and accuracy tests, confirms close agreement between silicon measurements, post-layout simulations, and software inference on the \emph{arrows8} dataset, with an average deviation of just 2.7\%, underscoring the reliability and real-world viability of the approach.

Measurements across multiple chips have confirmed the anticipated energy-efficiency potential of the ACNN chip. To maintain a consistent comparison framework, measurements at 30 operations were performed, showing that our chip achieves an average reduction of at least 2.1× in energy dissipation relative to an equivalent CMOS circuit, correlating with a predicted post-layout simulation reduction of approximately 2.8×. The experiments have further shown the capability of the chip to continue functioning correctly for a number of operations, without requiring the supply reservoir to be topped up, leading to a diminished supply voltage, which results in further energy efficiency. In one test, at 500 operations, the energy per synapse operation is reduced by an average of 6.8×.
This continuous operation without intermediate recharging of the tank capacitor and improved energy efficiency enables a more energy-stable regime, reducing cumulative losses and highlighting a unique operational advantage of the adiabatic architecture.

We have demonstrated that our ACNN chip delivers an effective combination of accurate classification and ultra-low energy consumption. Beyond these immediate performance metrics, a central contribution of this work is the deliberate way in which the circuit has been designed to support future, practical scaling. For example, we have already implemented multi-layer functionality and incorporated a bespoke routing layer that anticipates scenarios in which input (presynaptic) signals must travel long distances to reach their destination (postsynaptic) targets an essential capability for larger, more complex neuromorphic systems. Likewise, the experiments showing continuous, stable operation without refilling the PC supply tank were not conducted solely to highlight energy savings; they also demonstrate the system’s ability to tolerate a broad range of power-supply voltages, a critical requirement in real-world environments where operating conditions cannot always be tightly regulated. In addition, the ACNN layout was developed using a modular, tileable architecture, allowing individual blocks to be replicated and interconnected without redesigning core circuitry. This modularity is essential for scaling to much larger systems while maintaining consistent performance across tiles. We also evaluated the circuit’s behaviour under device mismatch conditions, confirming that the adiabatic operation and chosen neuron/synapse circuits remain stable even when subjected to parameter variation. Such robustness is a key requirement for manufacturable large-scale AI hardware 

Taken together, these architectural choices, experimental validations, and methodological advances underscore both the novelty and practicality of our adiabatic approach. They point to strong prospects for upscaling the concept and pave a realistic path toward deployment in operational settings. This is particularly relevant for power-constrained applications such as edge-AI platforms, UAVs, distributed sensing and infrastructure-monitoring systems, digital-twin environments, and other domains where energy efficiency and reliability are paramount.

\section{Methods}\label{sec:methods}

\subsection{Training of the ANN using TensorFlow}\label{sec:method_ann_training}
In our work, we used a two-layer, 64-input, 12-hidden neuron and 4-output software ANN. All neurons used 1-bit inputs and a single 1-bit output. The ANN was trained in Python (3.11.2) using TensorFlow~\cite{Abadi2016} (2.12.0) and Larq (0.13.0)~\cite{Geiger2020} AI frameworks. To ensure inference compatibility with the ACN architecture, all neurons used an ACN-compatible Heaviside activation function. However, during training, we used a \emph{tanh} activation function in the output layer neurons to aid training. This was then replaced with the Heaviside variant for inference, with only 0.5\% loss in overall accuracy. A classification rate of 98.65\% on the 4,078 independent \emph{arrows8} test samples was achieved after 200 training epochs using an Adam optimizer. Real-valued, signed weights were quantized to 8-bits using a modified Larq \emph{DoReFa} quantizer~\cite{Geiger2020} that constrained the weight values to an inclusive range of -1 to 1 with an extended dead-zone of $\pm 0.1$ around 0. This resulted in a 10:1 maximum-to-minimum weight ratio, giving us control over the mapped ACNs capacitance range. A fixed non-trainable software bias, $\tau = 0.1$, was used for all neurons.

\subsection{Mapping of ANN weights to ACN capacitance values}\label{sec:method_ann_mapping}
To ensure equivalent ANN to ACNN functionality, all of the real-valued, signed, AN weights, $w_{i}$, are converted to ACN capacitance values. This mapping was performed using the optimal \emph{conditional-mapping} process~\cite{Smart2025}, with
 \mbox{$V_{max}=1.5V$} used for the PC peak voltage and \mbox{$C_{min}=8fF$} for the minimum allowable capacitance value. The bias and ballast capacitive pillars, which provide additional capacitance values that do not affect computation, control the swing of the ACN membrane voltages, $v_{m}^{\pm}$. This ensured $v_{m}^{\pm}$ was always within the optimal [0.1-1]$V$ TL input voltage range. The total synapse capacitance, $C_{T}$, for each ACN was set as $C_{min}w_{T}/min |w_{i}|$, where $w_{T}=\sum{w_i}$ in each AN. As such, $C_{T}$ and, consequently, the physical size of each ACN in the chip, varied slightly.

The weight-to-capacitance value process was, theoretically, exact. However, there were two mapping-related practical considerations. The first was that the chip design layout used $\sim$~2$fF$ unit MOM capacitors, which introduced a small quantization error. The second, as described in~\cite{Smart2025}, was the top-plate parasitic-to-ground capacitance values on the $v_{m}^\pm$ nodes. A simple compensation process was used by removing the EDA tool-predicted node parasitic from the required ballast capacitance value. Approximately 8\% of the required ballast capacitance was derived from parasitics.

Extended Data Fig.~\ref{extfig:to2_config}a shows a breakdown of the ACN capacitance values. Some variations in ACN size within each layer were noted. Extended Data Fig.~\ref{extfig:to2_config}b shows a breakdown of capacitance value by type for the ACNN. Note that a large percentage of the ballast and bias was required simply for swing control. Extended Data Fig.~\ref{extfig:to2_config}c shows the breakdown of capacitance values, noting a large number of small weights coupled with a smaller number of dominant weights in each ACN. Finally, Extended Data Fig.~\ref{extfig:to2_config}d shows the distribution of quantization errors with a mean error of 0.53$fF$, which resulted in a small predicted loss of 0.39\% in overall classification.

\subsection{Internal configuration and control of the ACNN pipeline}\label{sec:method_control_logic}
Our ACNN chip is orchestrated by the control logic unit, shown in Fig.~\ref{fig:single_acn}b. The unit forwards the external PC signals ($PC1/2)$, and associated sampling clock signals ($CCLK1/2$) to the ACN and routing layers. Each sampling clock comprises of a peak clock ($CCLK_p$) for the DLCC and a trough clock ($CCLK_t$) for the latch. The power clock signal, $PC1$, drives all of our ACN capacitive synapses, whilst $CCLK1$ controls the 1-bit ACN TL sampling and output timing. The routing layer uses $PC2$, a second PC that is 180 degrees out of phase with $PC1$, in order to recover the charge from the adiabatic buffer. The clock signal, $CCLK2$, is used solely by the routing layer comparator and latch circuitry. 

During testing each 64-bit (8x8 1-bit pixel) test image was clocked in serially at $1MHz$ to the ACNN chip from an external SPI master, using standard SPI data signal ($DATA$), serial clock ($SCLK$) and chip select ($CS$). The active-low reset signal, $RST$, is a proprietary signal used to reset the $Deserializer$ logic ready for a new test image. The $Deserializer$ outputs a 64-bit parallel data bus, which acts as control signals to the 64 adiabatic switches in each first-layer ACN.

Finally, analog DC voltages $VBP$ and $VBN$ were routed from external pads to every ACN. During a reset period, when the signal $START=0$, Single Pole Single Throw (SPST) switches in each ACN are closed, allowing $VBP/VBN$  to be applied to the $v_{m}^{\pm}$ nodes. After $START$ goes high, these SPST switches are reopened, and computation can begin. The 12-bit output from the first layer, $QL1$, becomes valid after the first $PC1$ cycle. These bits then feed into the routing layer and, subsequently, layer 2, as well as being connected to external pads for debugging. The final computation, the image classification, becomes valid on $QL2$ after the second $PC1$ cycle, using a 4-bit one-hot encoding scheme.

\subsection{PCB PC clock generation design and operation}
\label{sec:PCG}
The Power Clock (PC) for our ACNN chip is generated on the PCB whose circuit is shown simplified in Extended Data Fig.~\ref{extfig:pcg}a. The circuit consists of a tank capacitor $C_{tank}=100nF$, inductor, $L_{pc}=0.39mH$, $C_{pc}=25pF$ and DC supply voltage, $V_{sup}=0.62V$. The switch $CHRG$ is used to connect the DC supply voltage to the tank capacitor, facilitating its charging phase. The tank capacitor can be charged at the beginning of each operational cycle by closing the switch ($CHRG$=logic high). Once the tank capacitor is charged the switch is opened to isolate the supply, thereby completing the charging sequence and preparing the system for the subsequent operational phase.

Following the charging phase, switch $PULSE$ is closed, connecting the tank capacitor to the inductor. This couples the tank capacitor with the inductor, initiating an LC resonance that allows energy exchange between the two components. This generates the required sinusoidal voltage at node $V_{pc}$. During this phase, switch $PULSEb$ remains open to avoid disrupting the oscillation. The duration for which switch $PULSE$ remains closed governs the frequency of the generated sinusoidal waveform. Once the oscillation completes, switch $PULSE$ is opened and switch $PULSEb$ is subsequently closed to reset the $V_{pc}$ node to ground, establishing the ideal phase spacing between successive sinusoidal cycles. The complete sequence of tank capacitor charging and the timed control of switches $PULSE$ and $PULSEb$ is illustrated in the simplified timing diagram shown in Extended Data Fig.~\ref{extfig:pcg}b.

Real PCG implementations, as implemented in our daughter board, include non-idealities, such as unwanted overshoot, reduced PC peaks, glitches and delays in TL clock generation signals, but also more subtlety, potential changes in frequency with capacitive load variations, as shown in Extended Data Fig.~\ref{extfig:pcg}c. A reduced maximum peak voltage of 1.47$V$ in $PC1$ was coupled with a 150$mV$ of unwanted undershoot. Small glitches and delays in the $CCLK$ generation derived from the PC are also possible. 

\subsection{Chip test configuration and custom ACNN daughter board}\label{sec:method_test_procedure}
Each of our ACNN chips was tested using the setup shown in the Extended Data Fig.~\ref{extfig:test_set_up}a. The system was controlled by software, connected to an ArC2 instrumentation board via a USB connection. Batched instructions for accurate timing were provided by the software to the ArC2 FPGA scheduler, which in turn drives the Device Under Test (DUT). Photographs of the test set-up are provided in Extended Data Fig.~\ref{extfig:test_set_up}b-c.

The ArC2 interfaces with a custom daughter board that hosts the DUT and generates the $PC$ signals using two LC ($L_{pc}=0.390mH$ and $C_{pc}=25pF$) oscillator circuits, as required by our ACNN chip. To replace lost charge during computation, each LC circuit was topped up from a charged $C_{tank}=100nF$ tank capacitor, acting effectively as a battery. The board also generates the associated $CCLK$ TL sampling signals for each PC, as previously discussed. 

The timing diagram in Extended Data Fig.~\ref{extfig:test_set_up}d shows a single computational operation. The ArC2 first generates a $10\mu s$ $CHRG$ strobe to charge both tank capacitors, up to the supply voltage, of around 0.62$V$. This tank capacitor can be topped up every operation, or just once, depending on the type of test. After raising the $RST$ signal, the 64-bit test \emph{arrows8} image is injected serially into the DUT via an SPI connection. Once loaded, the daughter board is instructed by the ArC2 to start generating the $PC$/$CCLK$ signals, 200$ns$ after the rising edge of $TRG$. A fixed number of sinusoidal-like peaks is then generated on each of the PCs: three on $PC1$ and four on $PC2$. After $START$ is raised, and the computation is completed,
software reads the 4-bit $QL2$ classification output.

\subsection{Measuring energy dissipation in the ACNN chip}\label{sec:method_measure_energy}
To measure the dissipated $PC1$ energy, we loaded an image into the ACNN deserializer, as normal. A multiple number of operations were then completed, \emph{with the same loaded input pattern}. This involved toggling the $TRG/START$ signals, but not $RST$, with a 1$\mu s$ idle-time in-between operations. Importantly, lowering the $START$ signal after each operation forces a reset of the ACN inputs, but not the deserializer. The latched deserializer contents could then be re-applied to the ACN inputs on the next rising $START$ edge, meaning the same amount of ACN switching per operation.

The second significant difference, compared to the functional classification tests, was that, after an initial charge, the two PCG tank capacitors were not topped up in between operations. As operations progressed, these tank capacitors would slowly drain, just like a battery. A Source Measure Unit (SMU) on the PCB monitored the voltage across each tank capacitor, $V_{tank}$, as shown in the Extended Data Fig.~\ref{extfig:pcg}a. The amount of stored energy in a tank, at any moment in time, was therefore determined by $0.5C_{tank}V_{tank}^{2}$. 
By monitoring the difference in energy before and after each multi-operation experiment, 
the total $PC1$ dissipated energy in the ACN layers, $E_{total}$, was measured.

A knock-on effect of draining the tank capacitors was that, as each experiment progressed, the $PC$ signals were subject to decay in their peak amplitudes over time.
This meant the computations were increasingly susceptible to miscalculation.
In this paper, the dissipated energy per synapse operation, $E_{SOP}$, was determined by
\begin{equation}
    E_{SOP}(O_{max}) = 
    \gamma
    \frac{E_{total}(O_{max}) - E_{PCG}(O_{max})}{N_{synapses} * O_{max} }
    \label{Eq: energy}
\end{equation}
where $N_{synapses}$ is the total number of ACNN synapses and $O_{max}$ is the maximum number of consecutive operations before the classification error rate exceeded 5\%. As we were interested only in the ACNN chip performance, we subtracted the expected energy loss from the PCB-based PCG, $E_{PCG}$, from the measured total, $E_{total}$. This baseline energy from the PCG was measured whilst operating without a DUT in place.
Finally, although three $PC1$ cycles are generated per operation, the computation was completed within two cycles. Consequently, a correction factor, $\gamma$, of $2/3$ was applied.

\backmatter

\bmhead{Acknowledgements}
This work has been, in part, funded by Defence Science and Technology Laboratory (Dstl), UK. We also acknowledge Professor Themis Prodromakis for kindly granting Dr Patrick Foster the opportunity to engage in this research. His support and willingness to foster collaboration were instrumental to the success of this work.




\newpage
\begin{appendices}

\section{Extended Data}\label{sec:extended_data}
\renewcommand{\thefigure}{\arabic{figure}}

\begin{figure}[h]
  \centering
  \includegraphics[width=\textwidth]{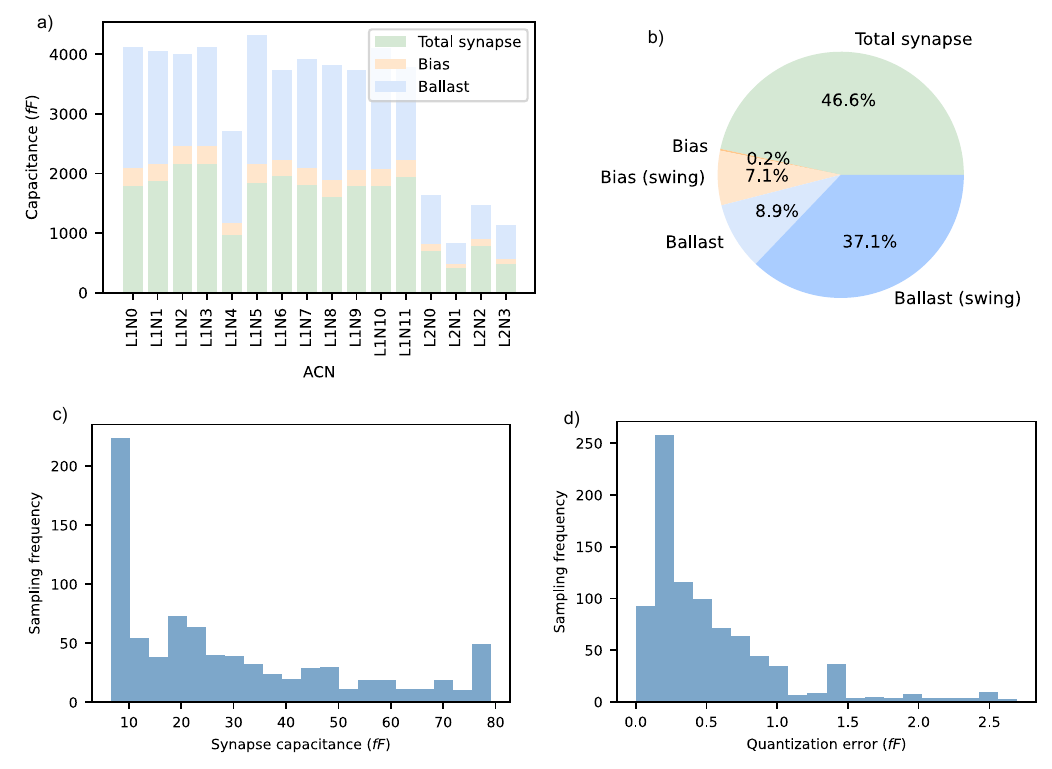}
    \caption{ACNN configuration: a) Capacitance value distribution across all 12 L1 ACNs and 4 $L2$ ACNs, showing contributions from the synapse, bias and ballast capacitors; b) Pie-chart showing contributions to the total ACNN capacitance of 51$pF$. Note, controlling the swing of membrane voltages to be constrained between 100$mV$ and 1.0$V$ is expensive; c) Distribution of synapse capacitance values between 8$fF$ and $80fF$; d) Distribution of quantization errors between all the required mapped 8-bit capacitance values and predicted post-layout capacitance values, with an average error of 0.53$fF$.}\label{extfig:to2_config}
\end{figure}

\clearpage
\begin{figure}[]
  \centering
  \includegraphics[width=\textwidth]{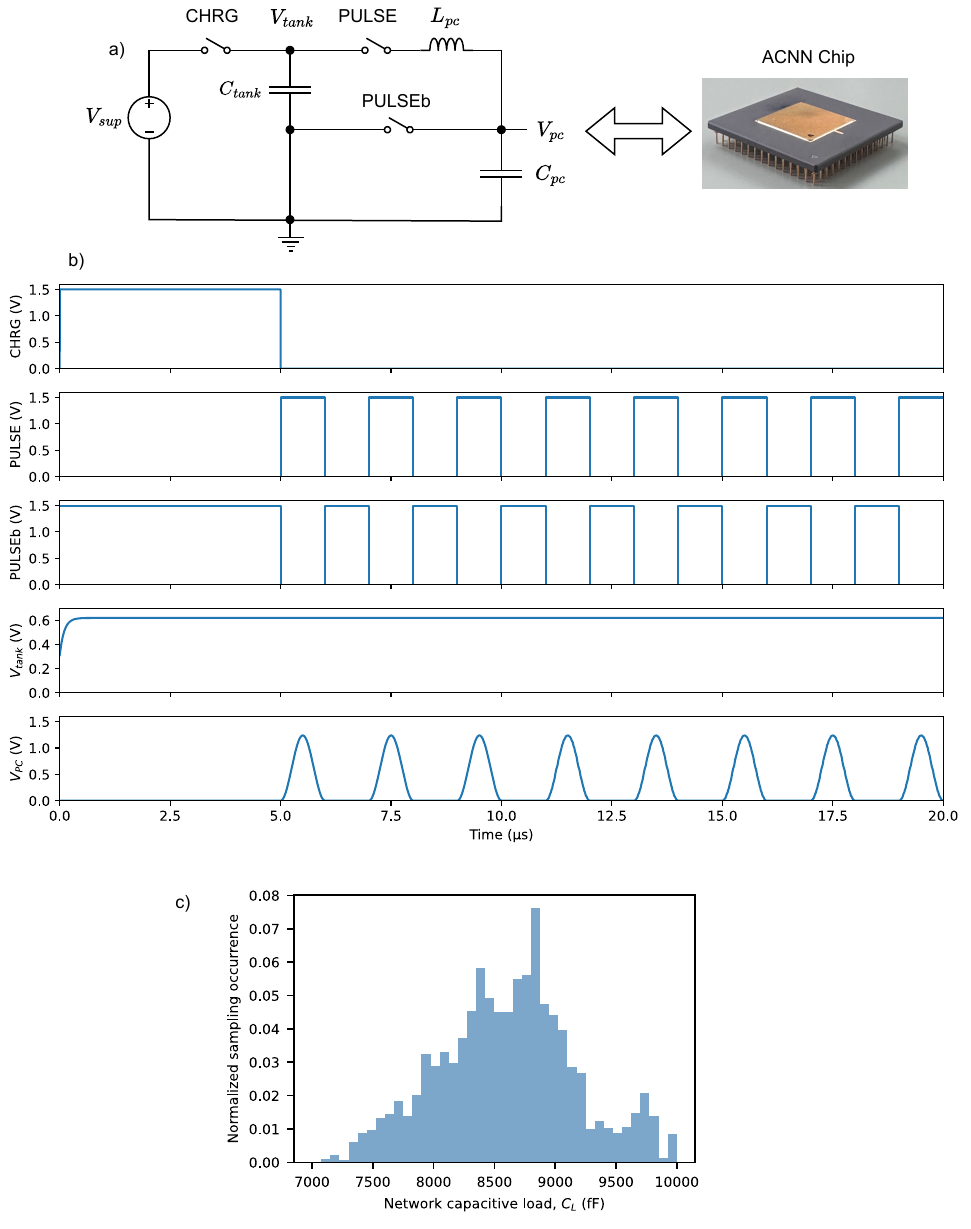}
    \caption{Power Clock Generator (PCG) design and operation: a) Daughter board power generator supply circuit design with tank capacitor, $C_{tank}=100nF$, inductor, $L_{pc}=0.39mH$, $C_{pc}=25pF$ and DC supply voltage, $V_{sup}=0.62V$; b) Timing diagram for a single capacitive tank charge, followed by a multi-cycle, PC ($V_{pc}$) operation with energy recovery, in and out of the ACNN chip; c) Software predicted variations in the network capacitive load, $C_{L}$, across the 4078 \emph{arrows8} test samples, with a mean $C_{L}$ of 8604$fJ$ and standard deviation of 765$fJ$. }\label{extfig:pcg}
\end{figure}

\clearpage
\begin{figure}[]
  \centering
  \includegraphics[width=\textwidth]{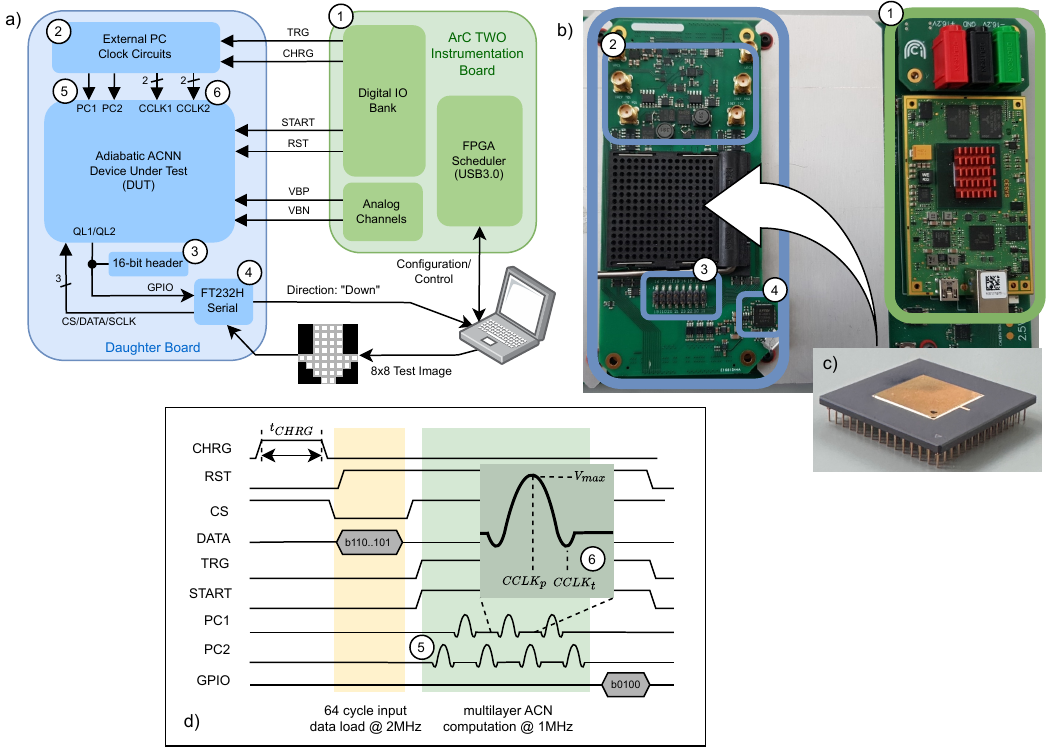}
    \caption{ACNN chip test setup: a) ArC2 and ACNN daughter board block diagram controlled via a laptop and two USB connections. Laptop configures the ArC2 to provide $VBN$ and $VBP$, typically 0$V$, analog voltages, accurately schedules the digital IO signals for PC clock generation circuits and configures/controls ACNN operations. The 64-bit test patterns are written to the ACNN deserializer via an FT232H serial chip; b) Photograph of the actual setup; c) A prototype ACNN test chip; d) Test system timing diagram for a single image classification operation. The PCB tank capacitor is charged first, followed by the loading of the 64-bit test data, before both the PC clock generators and ACNN are initiated by the $TRG$ and $START$ signals, respectively. After computation, the ACNN L2 classification outputs are read back to the software via the FT232H chip.}\label{extfig:test_set_up}
\end{figure}

\clearpage
\begin{figure}[]
  \centering
  \includegraphics[width=\textwidth]{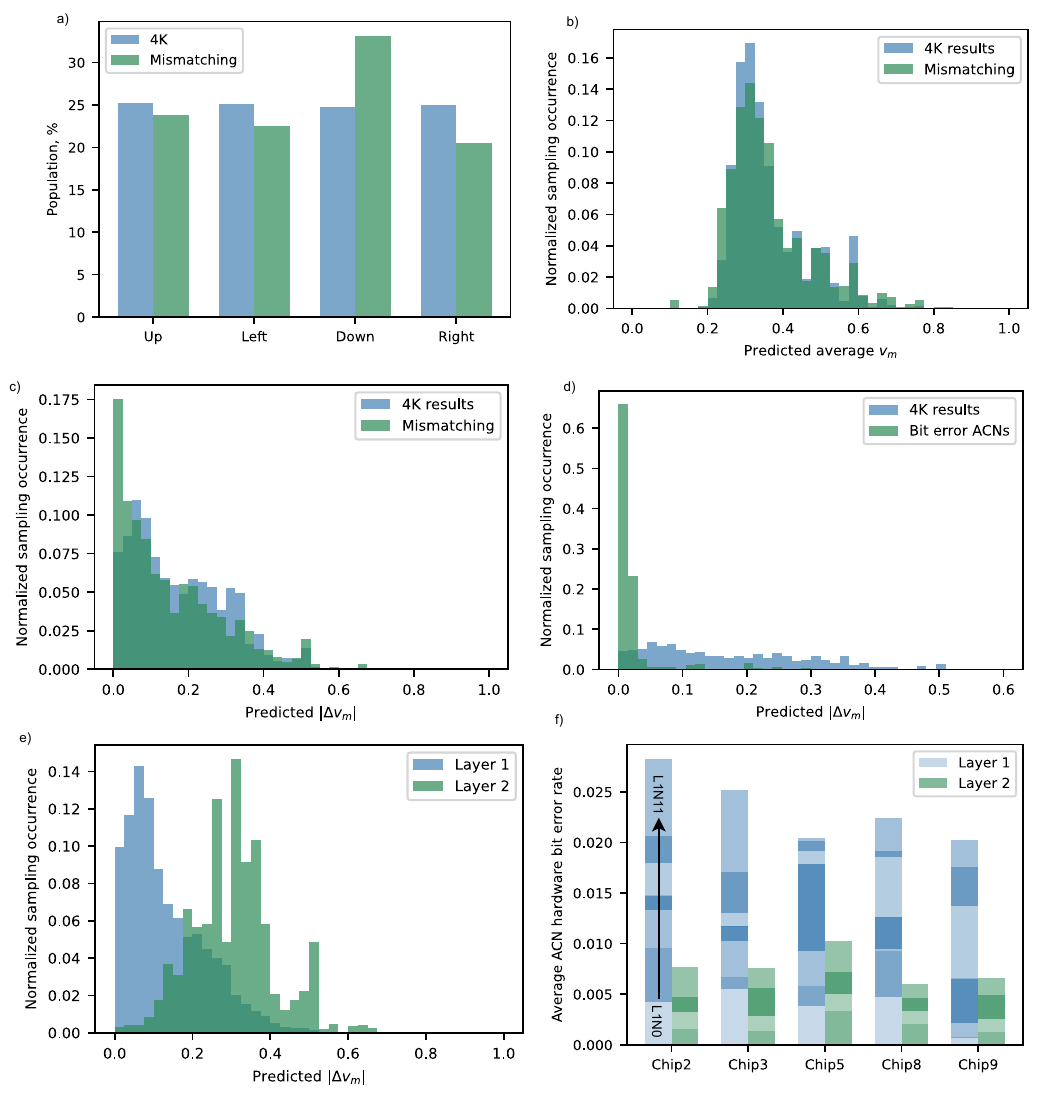}
    \caption{Analysis of the \emph{arrows8} classification results: a) class distribution of 4078 (4K) test samples versus the mismatching hardware sample (9.7\%) subset showing that \emph{Down} samples are slightly more likely to mismatch compared to other classes in the ACNN chips; b) Average ACN predicted membrane voltages, $v_{m}$, from software ACNN model, for all 4K test samples versus the mismatching subset; c) Full 4K dataset versus mismatching subset distributions for the predicted differential membrane voltages, $|\Delta v_{m}|$, across all ACNs; d) Full dataset versus mismatching subset distributions for the predicted $|\Delta v_{m}|$ for ACNs generating bit errors in the hardware; e) Predicted ACN $|\Delta v_{m}|$ across all $4K$ test samples split by layer; f) Breakdown of detected hardware bit errors across 5 chips and two layers with individual ACN bit errors stacked vertically with increasing neuron index. Note, only 8 layer  1 ACN outputs are monitored for bit errors due to board restrictions: $L1N0$ to $L1N5$ inclusive, $L1N8$ and $L1N11$.}
\label{extfig:to2_class_errors}
\end{figure}

\clearpage
\begin{figure}[]
  \centering
  \includegraphics[width=\textwidth]{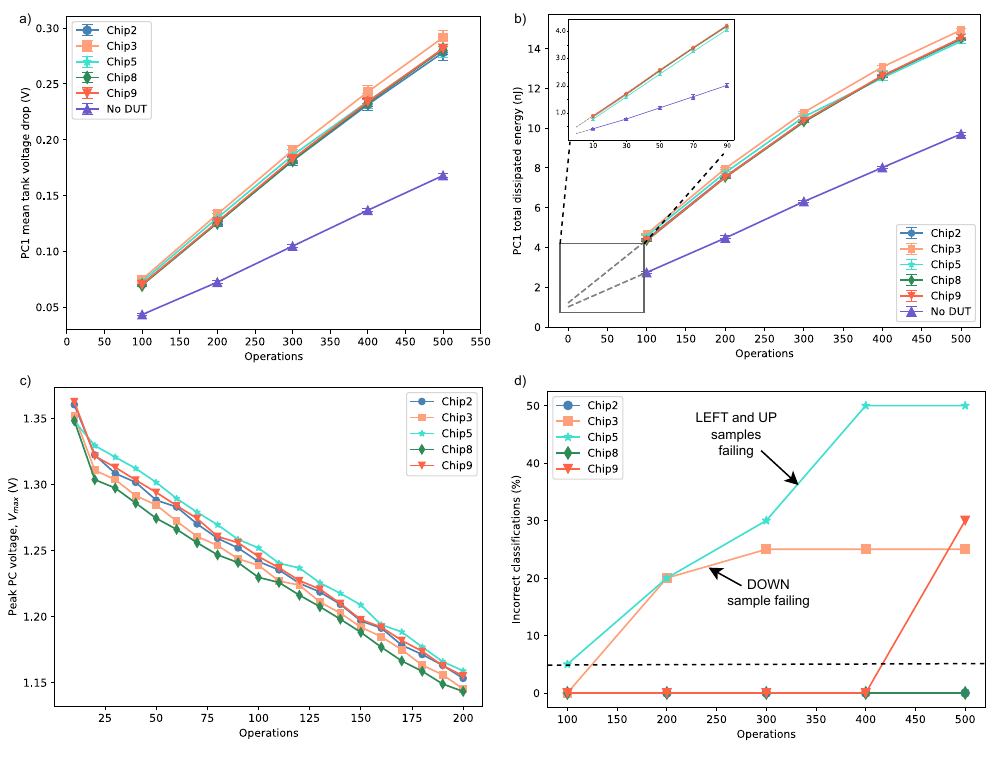}
    \caption{ACNN energy analysis: a) $PC1$ mean voltage drop across 100$nF$ tank capacitor, after a single charge, and followed by multiple operations of the same input, averaged across 4 test samples and 5 iterations per sample. Includes baseline case with no DUT; b) Total dissipated energy of $PC1$ of all ACNs under the same conditions;  c) Peak PC voltage, $V_{max}$, over number operations for the UP arrow; d) After several operations without a tank recharge, classification errors start and vary from chip-to-chip.}
\label{extfig:energy_analysis}
\end{figure}

\clearpage




\end{appendices}

\bibliography{sn-bibliography}

\begingroup
\renewcommand{\thefootnote}{}
\footnotetext{This work has been submitted to \emph{Nature Communications} for possible publication. Copyright may be transferred without notice, after which this version may no longer be accessible.}
\endgroup

\end{document}